To cite this paper, please use the following reference:

# Celeris: A GPU-accelerated open source software with a Boussinesq-type wave solver for real-time, interactive simulation and visualization

**Sasan Tavakkol, Patrick Lynett**

*Department of Civil and Environmental Engineering, University of Southern California, Los Angeles, California, USA*

**Abstract**

In this paper, we introduce an interactive coastal wave simulation and visualization software, called Celeris. Celeris is an open source software which needs minimum preparation to run on a Windows machine. The software solves the extended Boussinesq equations using a hybrid finite volume – finite difference method and supports moving shoreline boundaries. The simulation and visualization are performed on the GPU using Direct3D libraries, which enables the software to run faster than real-time. Celeris provides a first-of-its-kind interactive modeling platform for coastal wave applications and it supports simultaneous visualization with both photorealistic and colormapped rendering capabilities. We validate our software through comparison with three standard benchmarks for non-breaking and breaking waves.

**Keywords**: Wave Propagation; Wave visualization; Interactive modeling; Shader; Boussinesq; GPU

## 1 Introduction

Research with the Boussinesq-type equations has led to transformative changes in coastal engineering simulation and practice over the last few decades (e.g. [1]) These equations are powerful for the study of nearshore dynamics, including both nonlinear and dispersive effects. While Boussinesq-type equations are capable of simulating relatively short-waves, they are computationally more expensive than their counterpart, non-linear shallow water (NLSW) equations. The computational effort needed for Boussinesq-type equations hinders real-time simulations using them, requiring parallel processing on dozens to hundreds of CPU cores to achieve [2]. The needed supercomputing facilities are neither easily accessible nor inexpensive, particularly for the types of often low-budget coastal and civil engineering projects for which they are applicable. While Graphics Processing Units (GPUs), are affordable alternatives to accelerate these numerical models, they are not often leveraged for Boussinesq equations, perhaps because these equations do not easily lend themselves to highly parallel numerical schemes, due to their embedded implicit methods and large numerical stencils.

Finite volume method (FVM), shock-capturing, flux reconstruction, and limiters can make wave modeling solvers more robust; such approaches are now commonly found in NLSW models (e.g. [3]). However application of FVM to Boussinesq-type equations is not straightforward [4]. With FVM and the associated solution-smoothing schemes, robustness of the model becomes greater. This is of high relevance here, as our goal is to provide an interactive simulation environment, where the user can alter the water surface and the bathymetry while the model is running. This interactive environment also needs fast concurrent 3-D visualization. We choose a hybrid finite volume-finite difference scheme to solve the governing equations. This hybrid discretization enables the software to benefit from the robustness of FVM, while retaining the high accuracy of the Boussinesq-type model. To achieve high computational speed, we solve the equations using the GPU, providing faster than real-time simulation speed on an average user laptop. We call our open source software Celeris; the Latin word for "quick". The first version of our software is called Celeris Advent.

Celeris, to the best of our knowledge, is the first interactive software for simulation of nonlinear, coastal waves. In this software, the user can interact with the water surface and topography using system's mouse, for instance to add/remove water or raise/drop terrain. A GUI is also provided, by which the user can change the numerical and physical parameters on the fly. For example, a solitary wave can be added to the solution field or a sinewave can be introduced to a boundary, all while the model is running. The concurrent visualization in Celeris can also revolutionize the standard practice in the coastal engineering community. Currently, in well-known wave modeling codes, the simulation results are written to disk at certain checkpoints and are visualized using different tools afterwards. Simultaneous observation of results in Celeris, with photorealistic or colormapped rendering, can significantly help researchers to understand coastal processes in a specific event.

Our goal in development of Celeris was to provide a hassle-free software which can be run on off-the-shelf Windows machines with minimum preparation. Therefore we selected Microsoft's Direct3D library and its HLSL shader language to harness the power of the GPU. Since Direct3D is now included as an integral part of Windows operating systems, the compiled version of Celeris can be easily run on any recent Windows machine with a single click and without installation of any third-party software or library. Moreover, this implementation enables us to directly visualize the simulation results with the minimum overhead on the GPU and using Direct3D libraries. Celeris is implemented in C++ and HLSL, and it is an open-source code developed and redistributed under the terms of the GNU General Public License as published by the Free Software Foundation.

## 2   Mathematical model

### 2.1 Governing Equations

The extended Boussinesq equations derived by Madsen and Sørensen [5] are a suitable set for a hybrid finite volume-finite difference scheme [2]. These equations for 2DH flow read as:

$$\mathbf{U}_t + \mathbf{F}(\mathbf{U})_x + \mathbf{G}(\mathbf{U})_y + \mathbf{S}(\mathbf{U}) = 0 \tag{1}$$

$$\mathbf{U} = \begin{bmatrix} h \\ P \\ Q \end{bmatrix}, \mathbf{F}(\mathbf{U}) = \begin{bmatrix} P \\ \dfrac{P^2}{h} + \dfrac{gh^2}{2} \\ \dfrac{PQ}{h} \end{bmatrix}, \mathbf{G}(\mathbf{U}) = \begin{bmatrix} Q \\ \dfrac{PQ}{h} \\ \dfrac{Q^2}{h} + \dfrac{gh^2}{2} \end{bmatrix}, \mathbf{S}(\mathbf{U}) = \begin{bmatrix} 0 \\ ghz_x + \psi_1 + f_1 \\ ghz_y + \psi_2 + f_2 \end{bmatrix}$$

where $\mathbf{U}$ is the conservative variables vector, $\mathbf{F}(\mathbf{U})$ and $\mathbf{G}(\mathbf{U})$ are the advective flux vectors, and $\mathbf{S}(\mathbf{U})$ is the source term which includes bottom slope, friction, and dispersive terms. $h$ is the total water depth. $P$ and $Q$ are the depth-integrated mass fluxes in $x$ and $y$ directions respectively, where the $x$-$y$ plane makes the horizontal solution field. Subscripts $x$ and $y$ denote spatial differentiation, with respect to the corresponding direction, and subscript $t$ denotes temporal differentiation. $z$ is the bottom elevation measured from a fixed datum. $f_1$ and $f_2$ are the bottom friction terms and $g$ is the gravitational acceleration coefficient. $\psi_1$ and $\psi_2$ are the modified dispersive terms defined as:

$$\psi_1 = -\left(B + \frac{1}{3}\right)d^2(P_{xxt} + Q_{xyt}) - Bgd^3(\eta_{xxx} + \eta_{xyy})$$
$$- dd_x\left(\frac{1}{3}P_{xt} + \frac{1}{6}Q_{yt} + 2Bgd\eta_{xx} + Bgd\eta_{yy}\right) - dd_y\left(\frac{1}{6}Q_{xt} + Bgd\eta_{xy}\right) \tag{2}$$

$$\psi_2 = -\left(B + \frac{1}{3}\right)d^2(P_{xyt} + Q_{yyt}) - Bgd^3(\eta_{yyy} + \eta_{xxy})$$
$$- dd_x\left(\frac{1}{3}Q_{yt} + \frac{1}{6}P_{xt} + 2Bgd\eta_{yy} + Bgd\eta_{xx}\right) - dd_x\left(\frac{1}{6}P_{yt} + Bgd\eta_{xy}\right) \tag{3}$$

where $d$ is the still water depth and $B = 1/15$ is the calibration coefficient for dispersion properties of the equations. The free surface elevation is $\eta = w - w_s$, where $w$ is the water surface elevation and $w_s$ is the still water surface elevation both measured from the fixed datum. We use $[w, P, Q]^\top$ as the set of unknown variables, in which $w = h + z$. To avoid introducing unnecessary complication to the equations, we refrain from substituting $h$ with $w - z$; however, that is how $h$ is calculated in practice. Assuming constant bottom elevation in time, we have $w_t = h_t$.

The extended Boussinesq equations provide sufficiently accurate linear dispersion and shoaling characteristics for values of $kd < 3$, where $k$ is the wave number. Note that these equations automatically reduce to the Saint-Venant system of non-linear shallow water equations (NLSW) for $d = 0$. In locations where still water surface elevation is not defined, such as on lands above the sea level, we set $d = 0$ so the solver automatically switches to NLSW.

### 2.2 Numerical Model
Following Wei and Kirby [6], Eq. (1) can be rearranged as:

$$w_t = E(P, Q) \tag{4}$$

$$U_t^* = F(h, P, Q) + [F^*(Q)]_t \tag{5}$$
$$V_t^* = G(h, P, Q) + [G^*(P)]_t \tag{6}$$

where newly introduced quantities are defines by

$$U^* = P - \frac{1}{3}dd_xP_x - \left(B + \frac{1}{3}\right)d^2P_{xx} \tag{7}$$

$$V^* = Q - \frac{1}{3}dd_yQ_y - \left(B + \frac{1}{3}\right)d^2Q_{yy} \tag{8}$$

$$E(P, Q) = -(P_x + Q_y) \tag{9}$$

$$F(h, P, Q) = -\left(\frac{P^2}{h} + \frac{gh^2}{2}\right)_x - \left(\frac{PQ}{h}\right)_y - ghz_x - f_1 + Bgd^3(\eta_{xxx} + \eta_{xyy})$$
$$+ Bgd^2(d_x(2\eta_{xx} + \eta_{yy}) + d_y\eta_{xy}) \tag{10}$$

$$G(h, P, Q) = -\left(\frac{Q^2}{h} + \frac{gh^2}{2}\right)_y - \left(\frac{PQ}{h}\right)_x - ghz_y - f_2 + Bgd^3(\eta_{yyy} + \eta_{xxy})$$
$$+ Bgd^2(d_y(2\eta_{yy} + \eta_{xx}) + d_x\eta_{xy}) \tag{11}$$

$$F^*(Q) = \frac{1}{6}dd_xQ_y + \frac{1}{6}dd_yQ_x + \left(B + \frac{1}{3}\right)d^2Q_{xy} \tag{12}$$

$$G^*(Q) = \frac{1}{6}dd_xP_y + \frac{1}{6}dd_yP_x + \left(B + \frac{1}{3}\right)d^2P_{xy} \tag{13}$$

This rearrangement allows us to rewrite Eq. (1) as ODE's in time. The left hand side terms in Eq. (4)-(6) are discretized in time. $[F*(Q)]_t$ and $[G*(P)]_t$ are evaluated by extrapolation in time and the rest of the terms in the right hand side are known in the current time step. Following [2] and [4] we use a hybrid FVM-FDM discretization to solve these equations on uniform Cartesian grids. The spatial domain is discretized by rectangular cells with fixed dimensions of $\Delta x$ and $\Delta y$. Each cell plays the role of a control volume for the FVM discretization. Cell centers and their corresponding cell averaged values are used as the grid points in FDM. The advective terms along with the bottom slope term is discretized using a second-order well-balanced positivity preserving central-upwind scheme introduced by Kurganov and Petrova [3]. This scheme, sometimes known as KP07, is a finite volume method to solve the Saint-Venant system of shallow water equations. The rest of the terms are discretized using central FDM.

KP07 preserves stationary steady states (i.e. being well-balanced) and guarantees the positivity of the computed fluid depth. It supports a dry state with no need to keep track of the wet-dry front and it can accommodate discontinuous bottom topography. Moreover it is particularly suitable for implementation on the GPU [7]. We found this scheme to be a robust and accurate method, even with the single precision implementation of the GPU. The method is well-suited for interactive and high performance design of Celeris. Since the details of this scheme can be found in [3], we only describe its layout. The original KP07 for solving shallow water equations consists of these steps:

1- Unknown variables, $[w, P, Q]^\mathsf{T}$, are linearly reconstructed (evaluated) at cell interfaces applying a generalized minmod limiter on their derivatives.
2- A simple conservative correction is applied on $w$ to preserve the positivity of $h$. Flow velocities, $u$ and $v$, are calculated as

$$u = \frac{\sqrt{2}h(P)}{\sqrt{(h^4 + \max(h^4, \epsilon)}} , \qquad v = \frac{\sqrt{2}h(Q)}{\sqrt{(h^4 + \max(h^4, \epsilon)}} \tag{14}$$

where $\epsilon$ is a small predefined tolerance to avoid division by very small values or zero.
3- Fluxes are computed at each cell interface employing the central-upwind scheme.
4- Source terms are evaluated and unknown variables are found for the next time step.

In order to use KP07 as the FVM solver of our scheme, in its last step, we add the dispersive terms as source terms discretized by central FDM.

*2.3 Time integration*
Time integration is performed by a third-order Adams-Bashforth scheme as the predictor step, and an optional fourth-order Adams-Moulton scheme as the corrector step. The predictor step reads as

$$w_{ij}^{n+1} = w_{ij}^n + \frac{\Delta t}{12}\left(23E_{ij}^n - 16E_{ij}^{n-1} + 5E_{ij}^{n-2}\right) \tag{15}$$

$$U_{ij}^{*n+1} = U_{ij}^{*n} + \frac{\Delta t}{12}\left(23F_{ij}^n - 16F_{ij}^{n-1} + 5F_{ij}^{n-2}\right) + 2F_{ij}^{*n} - 3F_{ij}^{*n-1} + F_{ij}^{*n-2} \tag{16}$$

$$V_{ij}^{*n+1} = V_{ij}^{*n} + \frac{\Delta t}{12}\left(23G_{ij}^n - 16G_{ij}^{n-1} + 5G_{ij}^{n-2}\right) + 2G_{ij}^{*n} - 3G_{ij}^{*n-1} + G_{ij}^{*n-2} \tag{17}$$

where the superscripts denote the step number in time, with n being the last step with known variables. The predictor step is explicit in time, which means that all the variables on the right hand side of the equations are known. The corrector step is performed by

$$w_{ij}^{n+1} = w_{ij}^n + \frac{\Delta t}{24}\left(9E_{ij}^{n+1} + 19E_{ij}^n - 5E_{ij}^{n-1} + E_{ij}^{n-2}\right) \tag{18}$$

$$U_{ij}^{*n+1} = U_{ij}^{*n} + \frac{\Delta t}{24}\left(9F_{ij}^{n+1} + 19F_{ij}^n - 5F_{ij}^{n-1} + F_{ij}^{n-2}\right) + F_{ij}^{*p} - F_{ij}^{*n} \tag{19}$$

$$V_{ij}^{*n+1} = V_{ij}^{*n} + \frac{\Delta t}{24}\left(9G_{ij}^{n+1} + 19G_{ij}^n - 5G_{ij}^{n-1} + G_{ij}^{n-2}\right) + G_{ij}^{*p} - G_{ij}^{*n} \tag{20}$$

The corrector step is implicit in time. In order to solve it, the *n*+1 terms are calculated by the predictor step (or the corrector values from the previous corrector iteration) then the corrector step is iterated for a predefined number of times, or until the variables converge. Since the variables at previous time steps are not defined in the very first two time steps of the simulation (i.e. n=1 and n=2), a first order Euler time integration is used for those two steps.

The water surface elevation, $w^{n+1}$, is directly found by solving Eq. (15) or (18). However in order to calculate the flux terms, $P^{n+1}$ and $Q^{n+1}$ the following set of implicit equations must be solved:

$$A_{ij}^x P_{i-1,j} + B_{ij}^x P_{ij} + C_{ij}^x P_{i+1,j} = U_{ij}^* \tag{21}$$
$$A_{ij}^y Q_{i,j-1} + B_{ij}^y Q_{ij} + C_{ij}^y Q_{i,j+1} = V_{ij}^* \tag{22}$$
where

$$A^\alpha = \frac{dd_\alpha}{6\Delta\alpha} - \left(B + \frac{1}{3}\right)\frac{d^2}{\Delta\alpha^2}, \quad B^\alpha = 1 + 2\left(B + \frac{1}{3}\right)\frac{d^2}{\Delta\alpha^2}, \quad C^\alpha = -\frac{dd_\alpha}{6\Delta\alpha} - \left(B + \frac{1}{3}\right)\frac{d^2}{\Delta\alpha^2} \tag{23}$$

Eq. (21)/Eq. (22) results in a tridiagonal system of equations for each row/column of cells in the *x/y* direction. In order to efficiently solve these set of equations on the GPU, we use the cyclic reduction (CR) method which is described in more detail later.

## 2.4 Boundary conditions

Two layers of ghost cells are considered at each boundary and are used to implement the boundary conditions. Three types of boundary condition are implemented in Celeris Advent: sinewave maker, sponge layer, and fully reflective solid wall. More options will be available in the subsequent versions of Celeris, including random directional waves and boundary conditions set by time series input.

### 2.4.1 Solid wall

Solid walls are considered as fully reflective boundaries. In order to impose this condition the values on the closest two cells to the boundary are mirrored on the ghost cells. Mirroring ensures the following conditions are met:

$$(P, Q).\,\mathbf{n} = 0, \qquad \nabla w.\,\mathbf{n} = 0, \tag{24}$$
where **n** is the normal vector to the solid wall.

### 2.4.2 Sinewave maker

In order to generate sinewaves with a given period (*T*), amplitude (*a*), and direction (θ), at the boundary, the values for *η*, *P*, and *Q* are assigned as follows

$$\eta = a\sin\left(\omega t - k_x x - k_y y\right) \tag{25}$$
$$P = c\cos\theta\,\eta \tag{26}$$
$$Q = c\sin\theta\,\eta \tag{27}$$
where

$$c = \frac{\omega}{k}, \qquad \omega = \frac{2\pi}{T}, \qquad k_x = \cos(\theta)\,k, \qquad k_y = \sin(\theta)\,k \tag{28}$$

$k$, the wave number, is calculated using Eckart's [8] approximate solution for the dispersion relation:

$$k = \frac{\omega^2}{g} \sqrt{\coth\left(\frac{\omega^2 d}{g}\right)} \tag{29}$$

This implementation does not allow treatment of waves approaching the boundary and it can be used only if nonlinearity is insignificant.

### 2.4.3 Sponge layer

Sponge layers in Celeris are implemented following [4], by multiplying the values of $\eta$, $P$, and $Q$ by a damping coefficient defined by

$$\gamma(x,y) = f(x) = \frac{1}{2}\left(1 + \cos\left(\pi \frac{L_s - D(x,y)}{L_s}\right)\right) \tag{30}$$

where $L_s$ is the width of the sponge layer, and $D(x,y)$ is the normal distance to the absorbing boundary. Coefficient define by Eq. (30) is only applied to cells which are located inside the sponge layer.

### 2.5 Wave breaking

Wave breaking is not implemented in Celeris with a direct treatment. However, our experiments show that the numerical dissipation of the scheme caused primarily by using the minmod limiter imitates physical dissipation introduced by wave breaking. As discussed before, the solver to simulate the run-up on the beach automatically switches to the NLSW equations.

### 2.6 Friction

Friction terms in Eq. (1), which are particularly significant in run-up measurements, are given by:

$$\begin{bmatrix} f_1 \\ f_2 \end{bmatrix} = f \begin{bmatrix} P \\ Q \end{bmatrix} \frac{\sqrt{P^2 + Q^2}}{h^2} \tag{31}$$

where $f$ is the friction coefficient. In Celeris, the user can either opt to set the friction coefficient as a constant value or use the Manning's equation to derive it locally as:

$$f = \frac{gn^2}{h^{1/3}} \tag{32}$$

where n is the Manning's roughness coefficient. To avoid division by very small values of $h$ or zero, the same technique as in (14) is used.

### 2.7 Solitary waves

A solitary wave propagates on a horizontal bottom at a constant celerity and without change in its shape. Boussinesq equations permit such a wave with stationary shape provided that non-linear and dispersive effects are in balance. Celeris can take a set of solitary waves in its input file, with given wave heights, directions and crest locations. These waves can be also added later via the GUI and while the model is running. We superpose a solitary wave to the solution domain by adding $\eta$, $P$, and $Q$ in each cell by values given by:

$$\eta_s = H_s \operatorname{sech}\left(k_s\big((x - x_0)\cos(\theta) + (y - y_0)\sin(\theta)\big)\right)^2 \tag{33}$$

$$\begin{bmatrix} P_s \\ Q_s \end{bmatrix} = c_s \eta_s \begin{bmatrix} \cos(\theta) \\ \sin(\theta) \end{bmatrix} \tag{34}$$

where $H_s$ is the solitary wave height, θ is its direction, and ($x_0$, $y_0$) is the initial crest location. $k_s$ and $c_s$ are wavenumber and celerity of the solitary wave given by:

$$k_s = \sqrt{\frac{3|H_s|}{4d^3}} \qquad\qquad (35)$$

$$c_s = \sqrt{g(H_s + d)} \qquad\qquad (36)$$

Using the absolute value of $H_s$ in Eq. (35) allows insertion of a depression wave (i.e. single trough) in the software with negative wave heights. However, such a wave is not expected to maintain its shape.

## 3    Software Documentation

The fast computational speed of Celeris comes from its GPU implementation for solving the governing equations and visualizing the results. We distribute Celeris in its compiled version along with its open-source codes under GNU General Public License as published by the Free Software Foundation. We recommend users to work with the compiled version as much as possible, and try to recompile the software only if necessary. It must be added that shader files are compiled at runtime, therefore careful changes in those files do not require recompilation of the software. For instance, partially reflective boundary condition can be introduced by changing the code for sponge layer boundary condition in "compute.hlsl" without any recompilation.

### 3.1 Source files

Celeris is written in C++ and Microsoft's shader language, HLSL, and it is coded on top of an earlier open source demo project for modeling shallow water flows (Stephen Thompson, personal communication). Fig. 1 shows the simplified diagram of software flow in Celeris. The file named "main.cpp" takes care of the flow including reading the input file and calling appropriate functions in the loop. The bulk of the code is found in "engine.cpp". This file contains all the codes that drive the GPU and calls appropriate shaders for simulation and graphics rendering. It also writes data on disk at an optional user defined frequency. Simulation shaders are found in "compute.hlsl" and graphics shaders are in "graphics.fx". Finally the GUI is managed by "gui_manager.cpp".

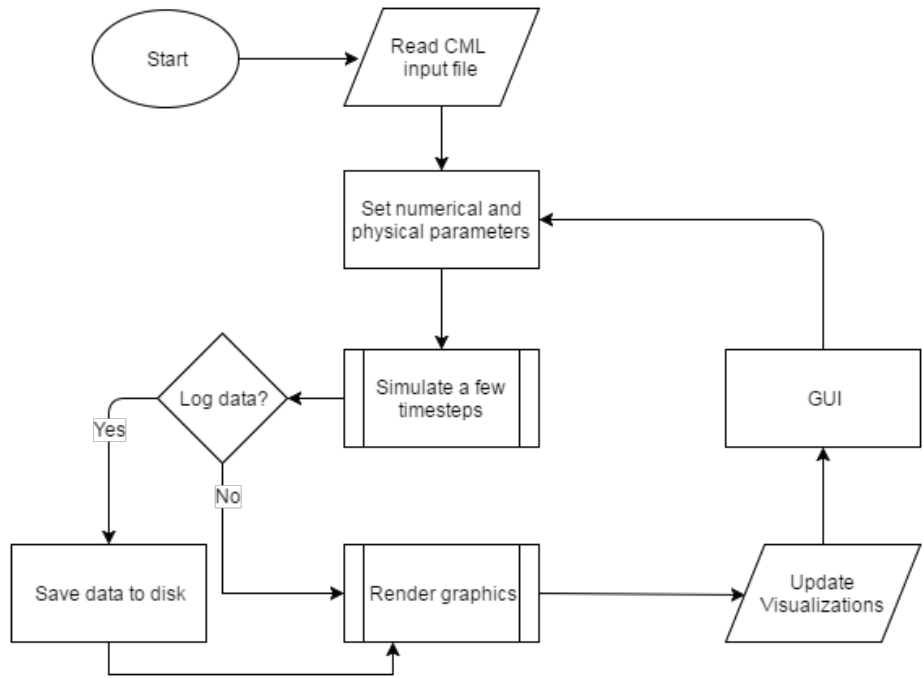

**Fig. 1. Simplified flowchart of Celeris.**

### 3.2 Input and output files

The input setup for a specific experiment can be given to Celeris as an XML (EXtensible Markup Language) file. XML files can be easily edited by any standard text editor. They are encoded with a set of labels (tags) in a format which is readable for both human and machine. In order to distinguish Celeris XML input files from generic XML files, we use CML as the format of these files. A sample CML file is shown in Fig. 2. In the input file, the model type can be chosen between Boussinesq and NLSW. The friction equations can be also selected to be Manning or Quadratic. The field dimension and grid sizes must be also entered in the input file. The bathymetry (topography) of the domain can be given as the relative or absolute path to a formatted ASCII file. The initial condition can be also set by entering the path to a formatted ASCII file which contains the initial values for $w$, $P$, and $Q$. Moreover, several solitary waves can be placed as the initial conditions. The boundary types must be chosen for each boundary among "Solid", "Sponge", and "SineWave". Most of the values given in the input file to the software can be later altered via GUI.

Finally the user can opt to save the $w$, $P$, and $Q$ data periodically on the disk at its associated cost. To minimize the slow-down, the user can choose to only save data on specific grid points (gauges) and/or several ranges. The output files are written in to a formatted ASCII file. In the next version of Celeris, we will add the option to write output files in a more efficient format, such as NetCDF.

```xml
<?xml version="1.0"?>
<Experiment>
    <name>Sample Experiment</name>
    <!-- Settings for Model -->
    <model type = "BSNQ">
        <parameters epsilon = 5e-12 correctionStepsNum = 2 timestep = 0.005></parameters>
        <friction type = "Manning" coef = 0.0> </friction>
    </model>
    <!-- Settings for Solution field -->
    <fieldDimensions width = 30 length = 30 stillWaterElevation = 0></fieldDimensions>
```

```
        <gridSize nx = 601 ny = 601></gridSize>
        <bathymetryFilePath> \resources\bathy.cbf </bathymetryFilePath>
        <!-- Settings for Initial Condition -->
        <hotStartFilePath> N/A </hotStartFilePath>
        <solitaryWave H = 0.05 theta = 0   xc = 5 yc = 15></solitaryWave>
        <solitaryWave H = 0.05 theta = -45 xc = 5 yc = 25></solitaryWave>
        <!-- Settings for Boundaries-->
        <westBoundary type = "SineWave" seaLevel = 0 widthNum = 2>
            <sineWave amplitude = .01 period = 2 theta = 0></sineWave>
        </westBoundary>
        <eastBoundary  type = "Sponge"  seaLevel = 0 widthNum = 20></eastBoundary>
        <southBoundary type = "Solid"   seaLevel = 0 widthNum = 2></southBoundary>
        <northBoundary type = "Solid"   seaLevel = 0 widthNum = 2></northBoundary>
        <!-- Settings for Logging Data-->
        <logData doLog = true logStep = 20>
            <logPath>C:\conical_island\</logPath>
            <range filename = "island">
                <bottomLeft x = 228 y = 228></bottomLeft>
                <topRight x = 374 y = 374></topRight>
            </range>
            <gauges filename = "gauges">229,302,249,302,353,302,354,302,301,249</gauges>
        </logData>
</Experiment>
```

**Fig. 2. Sample CML input file**

### 3.3 Implementation

Shader languages such as HLSL are designed around the idea that GPUs generate pictures [9]. Therefore, in order to solve a computational problem with shaders, the problem must be reformulated in terms of graphics primitives and the data must be stored within textures. 2D textures are matrix-like data structures which are well-suited for our 2D domain. Each cell in a texture, a texel, may have several floating point variables in order to describe traits of the texel. In Celeris, we mostly use float4 type texels which include three single precision floating point variable for texel color, namely "r", "g", "b", and one for the alpha channel, named "a". We use these variables to store flow parameters. For instance, a 2D-texture of size $n_x \times n_y$ is defined to store the latest state of the flow. In each computational cell, w, P, and Q are stored in "r", "g", and "b", while "a" is remained unused. Each step of numerical scheme described earlier is performed by passing several textures such as flow state, bathymetry, gradients, etc. as resources to a shader and getting one output, or as called in graphics terminology, render target texture. A sample shader to apply solid wall boundary condition is shown in Fig. 3.

```
float4 WestBoundarySolid(VS_OUTPUT input) : SV_TARGET
{
    const float3 in_state_real = txState.Load(int3(4 - input.tex_idx.x,input.tex_idx.y,0)).rgb;
    return float4(in_state_real.r, -in_state_real.g, in_state_real.b, 0);
}
```

**Fig. 3. Sample shader code which handles a solid wall boundary condition.**

After performing a user-defined number of computational time steps, the flow state and terrain are passed to the graphics renderer. Several shaders are applied in order to visualize the results with options for photorealistic rendering or value color-mapping.

The most challenging part of the implementation is solving the tridiagonal matrix systems within the numerical scheme. The classic algorithm to solve such a system is the Thomas algorithm consisting of a forward elimination and backward substitution. However this algorithm is inherently serial. Employing such an algorithm will generally need copying data from GPU to the main memory, running the serial solver and copying the results back on the GPU. Such a process will significantly increase the running time of the software and will become the bottle-neck for large domains. In Celeris, solving the tridiagonal system is accomplished using the cyclic reduction (CR) algorithm [10]. CR also consists of two phases: forward reduction and backward substitution. In the forward reduction phase, the system is successively reduced to a smaller system with half the number of unknowns, until a system of 2 unknowns is achieved which can be solved trivially. In the backward substitution phase, the other half of the unknowns are found by substituting the previously found values into the equations. This process is illustrated in Fig. 4

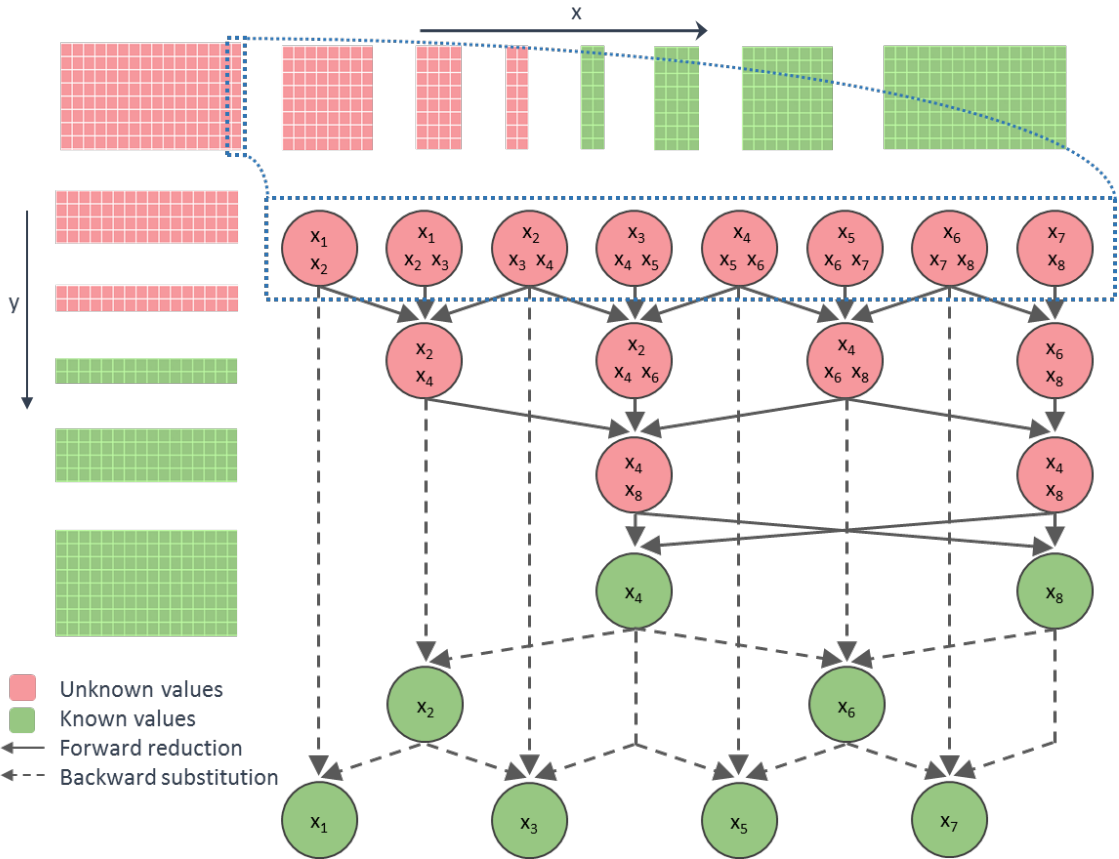

**Fig. 4. Cyclic reduction algorithm and its implementation on GPU.**

*3.4 Compilation*

Celeris is written and can be compiled in Microsoft Visual C++ 2008 Express Edition. The solution file named "Celeris.sln" is included in the redistributions. For successful compilation the latest DirectX SDK must be correctly installed. Celeris consists of three open source projects: a wrapper around operating system functions including Direct3D, named "Coerci", a GUI library named "Guichan", and the main project named "Celeris". The project Celeris uses an open source XML parser called "TinyXML". Two folders called "shaders" and "graphics" are also included in the redistribution zip file. These folders contains the shader codes and graphics textures (colormaps, font, etc.) and they must be placed appropriately in the solution folder such that they are found by the code.

*3.5 Running Celeris*

As mentioned in the previous sections, Celeris starts based on a CML input file; however the user can change most of the settings from the GUI while the model is running. Celeris can be easily launched by running the file named "Celeris.exe". After launch, the software will look into the file named

"setting.init" to find the absolute path to the input CML file. If such a path is not provided or the path is invalid, Celeris will ask the user to choose the input file from a file browser window.

After a successful launch, the numerical experiment begins immediately and the results are visualized in a 3D environment with a movable camera. Using the GUI, the user can change the numerical and physical parameters of the experiment such as the grid sizes, friction coefficient, boundary conditions, etc. Solitary waves can be also superposed to the field with a given location, height, and direction. Experiments can be paused or reset. The GUI of Celeris is briefly explained in a video available at https://youtu.be/pCcnPU7PCrg.

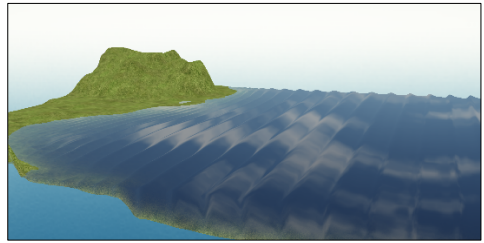

(a)

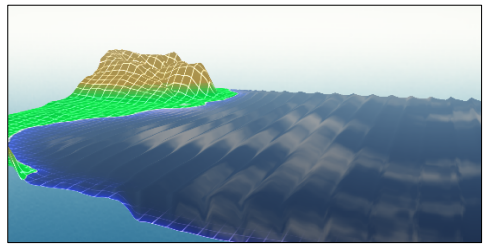

(b)

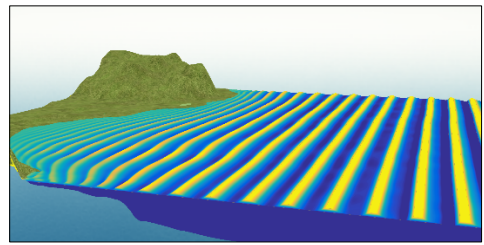

(c)

**Fig. 5. Visualizations of an experiment with a realistic bathymetry. Celeris can visualize the surfaces in a realistic mode with reflections and refractions (a), with a colormap on the terrain (b) and with a colormap on the water surface (c). Video is available at https://youtu.be/yks7ePXbRyU.**

Celeris provides various visualization options. The water surface can be visualized in a photorealistic mode, where reflection and refraction of rays hitting the water surface are calculated using the Fresnel equations, or by applying a colormap. This colormap can be set to represent $\eta$, $u$, $v$, or velocity magnitude of the flow. Several terrain textures are also available to enhance the visualization. The user can apply a colormap on the terrain as well. Finally a grid with a custom scale can be laid over the surfaces to improve the illustration of the surface elevation. Fig. 5 shows a combination of these different options for visualization of an experiment with a realistic relief and sinewaves on one of the boundaries.

## 4 Numerical validations

### 4.1 Run-up on a planar beach

Solitary wave propagation over a planar beach is experimentally studied by Synolakis [11]. In these experiments the beach slope was 1:19.85 and tens of trials were performed covering a wide range of solitary wave heights. This data set is used for numerical validation many times by several researchers [12], [13]. We simulate these experiments with a dozen wave heights in the range of 0.005 < H/d < 0.5 and we compare our numerical maximum vertical run-up to the experimental values. The chosen range for wave height covers both breaking and non-breaking waves. For simulations with H/d < 0.01 we use $\Delta x/h = 0.0625$ and $\Delta t\,(g/h)^{0.5} = 0.0075$ and for the rest of simulations we use $\Delta x/d = 0.25$ and $\Delta t\,(g/d)^{0.5} = 0.03$. The width of the simulation field is kept constant at W/d = 1 but the length is chosen between 100 < L/d < 1000 m such that it appropriately accommodates the solitary wave. The beach is located close to the east boundary. The west boundary is a sponge layer and the two other boundaries are solid walls. Following Lynett et al. [12], we generated three sets of experiments with different constant quadratic bottom friction coefficients, f = 0.0, 0.01, and 0.001. Fig. 6 compares the numerical results with experimental data, where maximum vertical run-up and solitary wave height are scaled by the water depth. For non-breaking solitary waves with H/d < 0.01, the bottom friction does not affect the maximum run-up, and the results agree quite well with experiments. For larger breaking waves, the numerical results for different bottom frictions begin to diverge. Note that Celeris does not employ an explicit wave-breaking model. However the minmod limiter used in the numerical scheme, introduces sufficient numerical dissipation to resemble wave breaking. The achieved results are consistent with results of Lynett et al. [12].

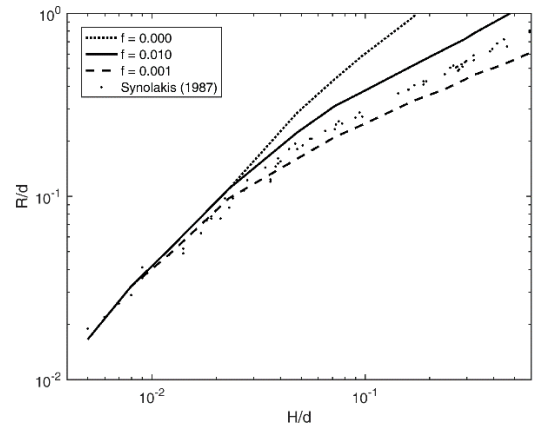

**Fig. 6. Comparison of numerical non-dimensional maximum run-up of solitary waves on a 1:19.85 beach versus non-dimensional wave height with experimental data**[11]**.**

Synolakis [14] also provides snapshots of the water surface elevation using photographs of the waves during the run-up and run-down. One particular set of these snapshots with H/d = 0.28 is used by several researches to evaluate their models. The results for this numerical experiment in Celeris is compared with experimental data in Fig. 7. Following [13] we used a friction factor of f = 0.0075 in this simulation. The comparisons indicate the ability of Celeris to accurately predict the run-up and run-down process for a breaking wave.

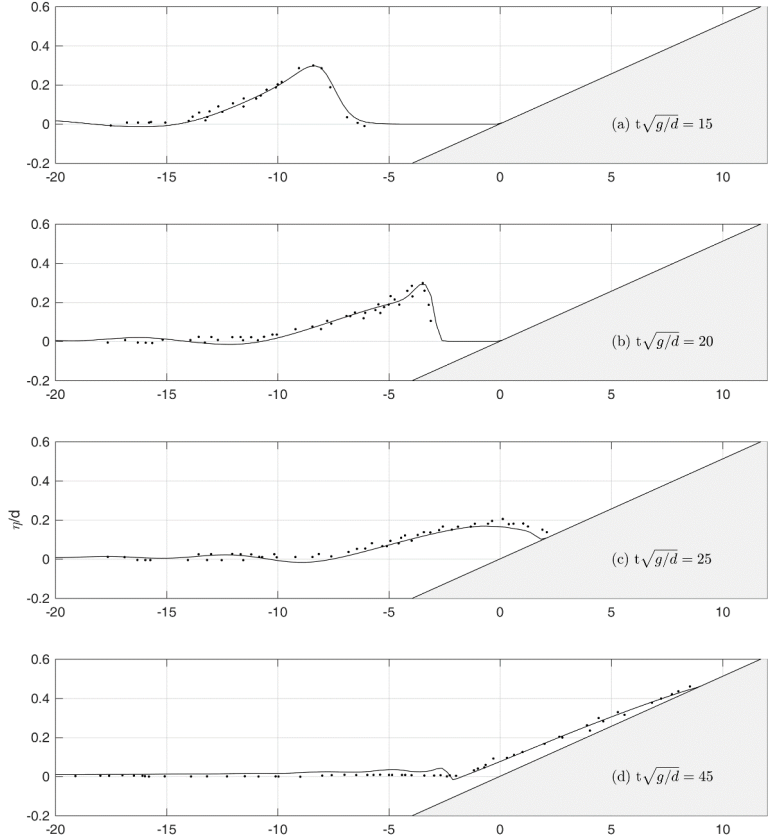

**Fig. 7. Breaking solitary wave run-up and rundown on a planar beach at t( g/h)^0.5 = (a) 15, (b) 20, (c) 25, (d) 45. The solid line represents the numerical results and the dots represent the experimental data of Synolakis** [14]

### 4.2 Wave focusing on a semicircular shoal

We extend our validation to 2-D problems by firstly simulating the experiments of Whalin [15]. He studied the non-linear refraction–diffraction of regular waves propagating over a semicircular shoal in a wave tank which was 25.6 m long and 6.096 m wide. The water depth in the tank was gradually decreased from 0.4572 m to 0.1524 m. The bathymetry can be expressed by

$$z = \begin{cases} 0 & 0 \leq x < 10.67 - G \\ (10.67 - G - x)/25 & 10.67 - G \leq x < 18.29 - G \\ 0.3048 & 18.29 - G \leq x \end{cases} \tag{37}$$

where $G(y) = [y(6.096 - y)]^{1/2}$, $0 \leq y \leq 6.096$. Harmonic analysis was performed on surface elevation time series along the tank centerline to obtain the amplitude of frequency components. The Whalin [15] experiments have become one of the standard benchmarks for Boussinesq wave models, and are used for model validation by several authors in previous studies [4], [16]–[18]. We study the case with incoming wave amplitude of a = 0.0075 m and period of T = 2 s.

We simulate this experiment with Celeris in 35 m x 6.096 m, imposing a sinewave boundary condition on the west boundary, and a 5 m sponge layer on the east boundary. The north and south boundaries are solid walls. The domain is discretized by 2000 x 65 cells with a time step of 0.001 s. On each cell along the centerline of the tank, the amplitudes of the first, second, and third harmonics are calculated based on FFT analyses and then they are compared to those of Whalin's experimental data in Fig. 8.

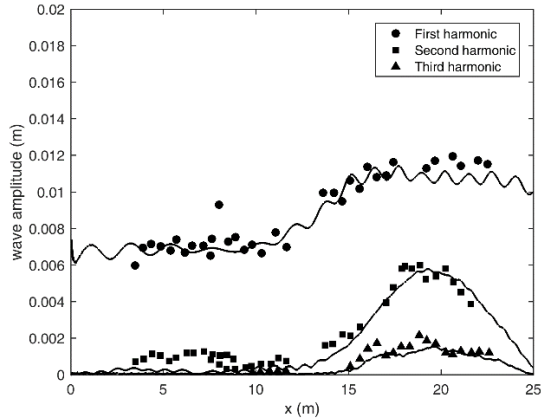

**Fig. 8. Wave amplitude harmonics along the centerline for $a$ = 0.0075 m and $T$ = 2 s. Solid lines are numerical results from Celeris, symbols are experimental data from Whalin [15].**

A snapshot of water elevation is shown in Fig. 9. The regular sinewaves coming from the boundary focus on the semicircular shoal, and higher harmonics appear due to the non-linear effects. The focusing of the waves can be clearly seen in Fig. 9. The vertical scale is exaggerated by a factor of 80 in the software.

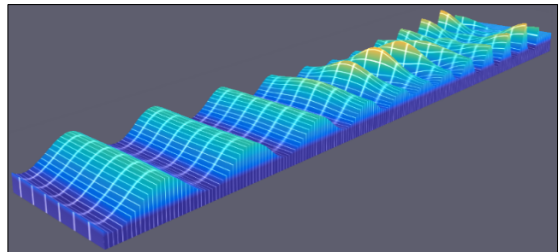

**Fig. 9. Water surface elevation for the case with $a$ = 0.0075 m and $T$ = 2 s from Whalin [15] experiments.**

*4.3  Solitary wave run-up on a conical island*

As the final validation test, we reproduce the experiments of Briggs et al. [19] for solitary wave interaction around a conical island; a test case frequently used to validate numerical models [12], [13], [20], [21]. The experimental setup is shown in Fig. 10. A circular island with 7.2 m base diameter and ¼ side slope was located in a 30 m x 25 m wave tank with 0.32 m depth. Three cases with target relative wave heights of $H/d$=0.05, 0.10, and 0.20 were simulated and the wave maximum run-up on the island and surface elevation time series on several gauges were recorded.

We simulate the conical island experiments in a 30 m x 30 m numerical domain with the conical island in the center and a soliton placed as an initial condition near the west boundary. Sponge layers are imposed on the boundaries parallel to the soliton, and solid walls on the two other boundaries. The domain is discretized by 601x601 cells with a constant time step of 0.005 s. Bottom friction is neglected in these simulations. The test cases are performed with relative wave heights of $H/d$=0.04, 0.09, and 0.18 which are slightly smaller than the target wave heights, but closer to those observed downstream of the wave maker. Reduced wave heights are also used by [12], [13], and [21].

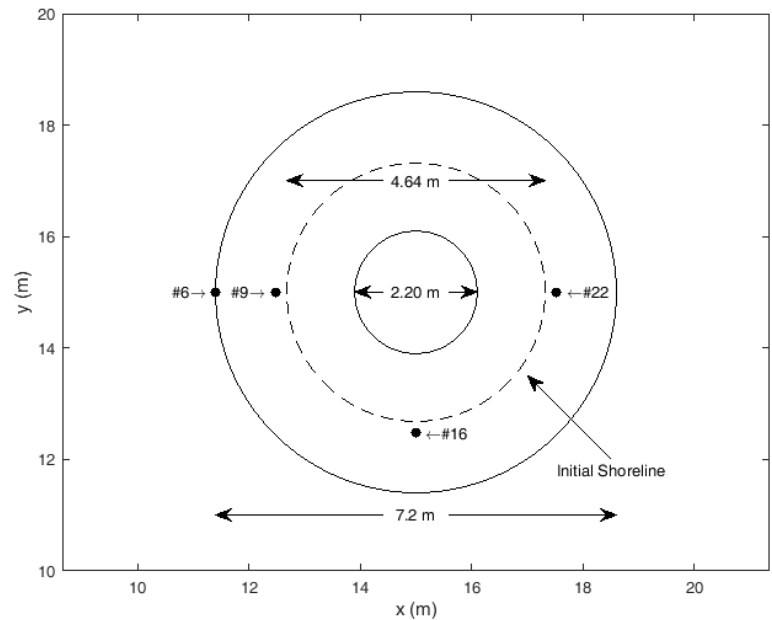

**Fig. 10. Experimental setup of the conical island. The gauge locations are shown by dots and the wave approaches the island from the left.**

Gauge #6 and #9 are located in front of the island, while gauge #16 is on the side, and gauge #22 is behind the island (Fig. 11). The numerical surface elevation compared to the experimental results are shown in Fig. 11 to Fig. 13. The leading wave height and its shape is predicted very well in all cases. The initial draw-down is also predicted quite well for two cases with larger wave heights. However the draw-down is underestimated for the case with the smallest wave height. This deviation is consistent with numerical results of the previously cited references.

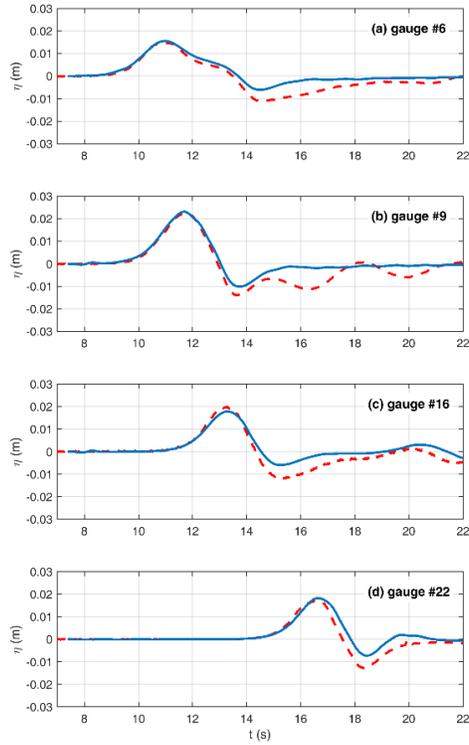

Fig. 11. Experimental (– –) and numerical (–) time series for the interaction of a solitary wave with H/d=0.04 on a conical island, at gauges #6, #9, #16, and #22 (a-d)

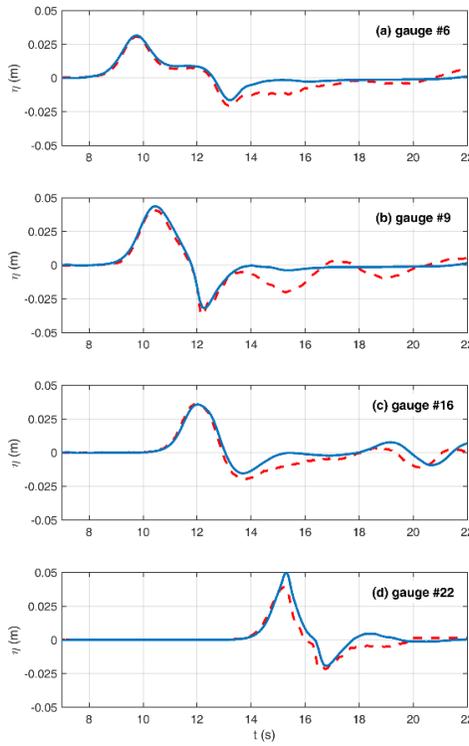

Fig. 12. Experimental (– –) and numerical (–) time series for the interaction of a solitary wave with H/d=0.09 on a conical island, at gauges #6, #9, #16, and #22 (a-d)

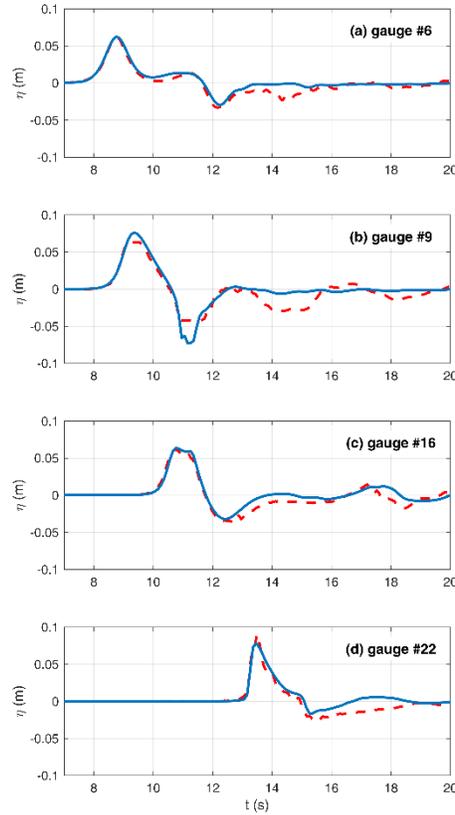

**Fig. 13.** Experimental (− −) and numerical (−) time series for the interaction of a solitary wave with H/d=0.18 on a conical island, at gauges #6, #9, #16, and #22 (a-d)

Snapshots of the experiment for the case with $H/d$=0.18 are shown in Fig. 14. The moment of maximum run-up on the front face of the island is shown in Fig. 14a. The time when the wrapping waves collide behind the island is captured in Fig. 14b. In these two figures, the water surface is rendered by a colormap representing the lateral velocity, v, where lighter colors represent positive values and darker colors represent negative values. Fig. 14c shows the time of the maximum run-up on the back face of the island. The colormap in this figure represents $\eta$.

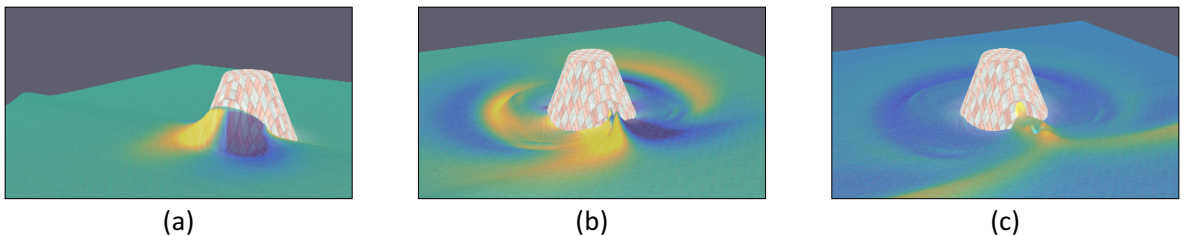

(a)                              (b)                              (c)

**Fig. 14.** Snapshots of conical island with $H/d$ = 0.18 near the time of maximum run-up at the front face (a), collision of wrapping waves (b), and maximum run-up on the back face (c). The vertical scale is exaggerated by a factor of 10

The predictions of the current model are slightly better than the numerical results in [12], [13], and [21]. For instance, the double-peak in Fig. 13c is resolved better in the current model. This might be because of the finer resolutions used in the current study, which were feasible only due to the fast computational speed of Celeris. For example, Fuhrman and Madsen [21] reported a 3.3 h simulation running time on a single 3.2 GHz Pentium 4 processor, with a 234×201 computational grid. Celeris completes this test with the same number of cells in less than 15 s, on a PC with NVIDIA Quadro K600 graphics card and a 1.8 GHz Intel Xeon CPU.

The agreement of numerical results with measured values in the case with the highest wave height is the most interesting one, as in this case the soliton breaks along the island. Fuhrman and Madsen [21] model, which does not utilize a breaking model, over predicts the run-up for gauge #22 by about 25% for this case. However predictions of Lynett et al. [12] and Tonelli and Petti [13], which consider the wave breaking, are much closer to the measurements. Our model also has a close prediction at this location, which confirms that the minmod flux limiter employed in Celeris is doing a good job in imitating the breaking models.

Finally, the numerical and measured horizontal maximum run-ups are compared in Fig. 15. The horizontal run-ups are scaled by the initial shoreline radius (2.32 m). The wave direction is from west to east. A threshold of $\delta = s\Delta x/3$ is chosen for water depth to determine the maximum run-up. The run-up values for the selected $\delta$ were invariant for different grid sizes. The agreement for all cases is very good and comparable to that achieved by Lynett et al.[12], Fuhrman and Madsen [21], and Tonelli and Petti [13]. The run-up on the back face of the island is also captured very well in these simulations. This run-up is generated by the collision of waves wrapping around the island.

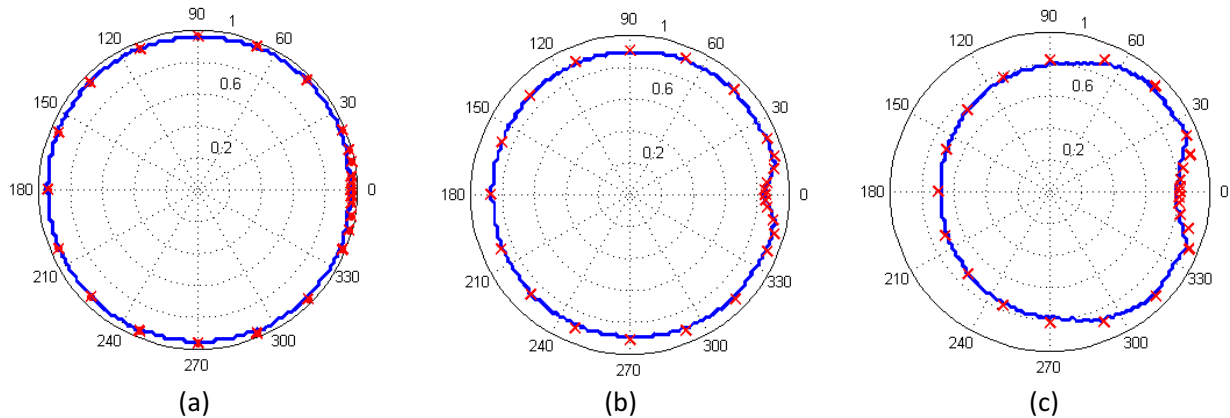

**Fig. 15. Numerical (solid line) and measured (x) maximum horizontal run-up for H/d = 0.04 (a), 0.09 (b), and 0.018 (c).**

# 5 Conclusion

An open source software for coastal wave simulation and visualization, called Celeris, is introduced. The discretization of the extended Boussinesq equations by a hybrid finite volume – finite difference scheme is briefly explained and its implementation on GPU is discussed. The structure of the software is sketched and its components are elaborated. Celeris is validated for breaking and non-breaking waves by comparing its results with three standard benchmarks; namely, run-up on a planar beach, wave focusing on a semicircular shoal, and solitary wave run-up on a conical island.

The main feature of the software, in addition to its fast computational speed, is its interactivity. The user can change the physical and numerical parameters of an experiment via a GUI, while the model is running. Numerous visualization options including photorealistic rendering are provided. A compiled version of Celeris is distributed along with its source codes under terms of the GNU General Public

License. Celeris harnesses the GPU by using Direct3D libraries, and it can run on any recent Windows machine with minimum preparation.

## Acknowledgements

The authors acknowledge Dr. Stephen Thomson, who developed the initial version of the solver and GUI, on which the Celeris software has been built.  Research support for this effort was provided by the Office of Naval Research.

Pasadena, CA. 1986.